\documentclass[onecolumn,showpacs,nobibnotes]{revtex4}

\usepackage{amsmath}
\usepackage{graphicx}

\begin{document}

\title{New insight in global QCD fits using Regge theory}
\pacs{11.55.-m, 13.60.-r}
\author{G. Soyez\footnote{FNRS Postdoctoral Researcher (Charg\'e de Recherches). On leave from the fundamental theoretical physics group of the University of Li\`ege.}}
\email{gsoyez@spht.saclay.cea.fr}
\affiliation{CEA Saclay, Service de Physique Th\'eorique, Orme des Merisiers B\^at 774, F-91191 Gif-Sur-Yvette, France}

%%%%%%%%%%%%%%%%%%%%%%%%%%%%%%%%%%%%%%%%%%%%
%% FRONTMATTER
%%%%%%%%%%%%%%%%%%%%%%%%%%%%%%%%%%%%%%%%%%%%

\begin{abstract}

In global QCD fits, one has to choose an initial parton distribution at $Q^2=Q_0^2$. I shall argue that the initial condition choses in usual standard sets is inconsistent with analytic $S$-matrix theory. I shall show how one can combine these two approaches, leading to a Regge-compatible next-to-leading order global QCD fit. This allows one to extend the parametrisation in the low-$Q^2$ region. Finally, I shall discuss how it it possible to use the Dokshitzer-Gribov-Lipatov-Altarelli-Parisi (DGLAP) equation to obtain information on Regge models at high $Q^2$.

\end{abstract}

%%%%%%%%%%%%%%%%%%%%%%%%%%%%%%%%%%%%%%%%%%%%
%% MAINMATTER
%%%%%%%%%%%%%%%%%%%%%%%%%%%%%%%%%%%%%%%%%%%%

\maketitle

\section{Introduction}

One of the most important results of QCD is the Dokshitzer-Gribov-Lipatov-Altarelli-Parisi (DGLAP) evolution equation \cite{DGLAP} which gives the evolution of the parton densities with the virtuality $Q^2$ of the photon. The kernel of these equation, {\em i.e.} the splitting functions, has been computed recently \cite{Moch:2004pa,Vogt:2004mw} at the three-loop level in perturbation theory (NNLO). From the phenomenological point of view, the large number of experimental data coming from the highly-accurate measurement of the structure functions, {\em e.g.} at HERA \cite{H1,ZEUS}, allows for the determination of the parton densities. 

The technique is to choose a parametric distribution for the parton at an initial scale $Q_0^2$, to obtain the Parton Distribution Functions (PDF) at all $Q^2$ using the DGLAP equation and to construct the physical quantities by convolution with the coefficient functions. One can then adjust the initial condition in order to reproduce the data as well as possible. This way of getting a standard PDF set, usually called {\em a global QCD fit}, has been applied many times with many updates by various teams \cite{CTEQ,MRST,GRV,ZEUSg,Alekhin}. 

Beside this approach, it is well-known that one can reproduce the hadronic data using the analyticity properties of the $S$ matrix. In this framework, the amplitudes are considered not as a function of the energy but in complex angular-momentum space. At high energy, the behaviour of the amplitude is then given by the leading singularities in the complex-$j$ plane, corresponding to pomeron and reggeons exchanges. Therefore, in order to reproduce the data, one chooses a structure of singularity in the complex-$j$ plane and adjusts their residues. Usually, we consider a pomeron contribution reproducing the growth of the cross-section at high energies, together with reggeon terms taking into account $f$- ,$a$- and $\omega$-meson trajectories exchanges. Within this framework, different models of pomeron are consistent with the present data \cite{compete}. As an example, one can consider the Donnachie-Landshoff two-pomeron model \cite{DL} with one simple pole at $j=1.4$ ($F_2\propto x^{-0.4}$) and a simple pole at $j=1.08$ ($F_2\propto x^{-0.08}$). Throughout these proceedings, we shall concentrate on another choice known as the triple-pole pomeron model \cite{triple}, where the pomeron singularity is a triple pole at $j=1$, corresponding to a cross-section growing like $\log^2(1/x)$.

\section{Regge-compatible global fit}

\subsection{The initial condition problem}

The DGLAP equation, being a renormalisation group equation, gives you the the $Q^2$-evolution of the PDF. This means that you still have to provide an initial condition at an initial scale $Q_0^2$. Let us consider, {\em e.g.}, the CTEQ 6M \cite{CTEQ} initial distributions. At small $x$, they obtain
\[
xq(x) \propto x^{-0.30}\quad\text{ and }\qquad xg(x)\propto x^{0.51}.
\]
This result suffers from two problems: first, the corresponding $j$-plane singularities are not seen in soft amplitudes. Secondly, quarks and gluons, while being coupled, do not have the same singularity structure. These two problems are in contradiction with the fact that Regge theory requires \cite{Cudell:2002ej} all hadronic amplitudes to have the same $j$-plane singularities.

\subsection{Proposed solution}
\begin{table}
\begin{tabular}{lccccccc}
\hline
\multicolumn{2}{c}{Data} & \multicolumn{3}{c}{LO global fit} & \multicolumn{3}{c}{NLO global fit}\\
\hline
%\hline
Exp.          & nop  & $Q^2\ge 5$ & $Q^2<5$ & total & $Q^2\ge 5$ & $Q^2<5$ & total \\
\hline
$F_2^p$       & 1828 & 0.969 & 0.867 & 0.948 & 0.943 & 0.933 & 0.937 \\
$F_2^d$       &  391 & 1.075 & 1.124 & 1.089 & 1.073 & 1.041 & 1.064 \\
$F_2^n/F_2^p$ &  211 & 1.233 & 0.886 & 1.051 & 1.270 & 0.764 & 1.004 \\
$F_2^\nu$     &   84 & 2.385 & 5.063 & 2.767 & 2.096 & 4.353 & 2.418 \\
$xF_3^\nu$    &  111 & 0.440 & 1.420 & 0.652 & 0.670 & 0.613 & 0.658 \\
\hline
Total         & 2615 & 1.026 & 1.017 & {\bf 1.023} & 1.008 &0.975 & {\bf 1.000} \\
\hline
\end{tabular}
\caption{Result of the global QCD fit together with the Regge small-$Q^2$ extension.}\label{tab:chi2}
\end{table}

\begin{figure}
\includegraphics[height=.32\textwidth,angle=270]{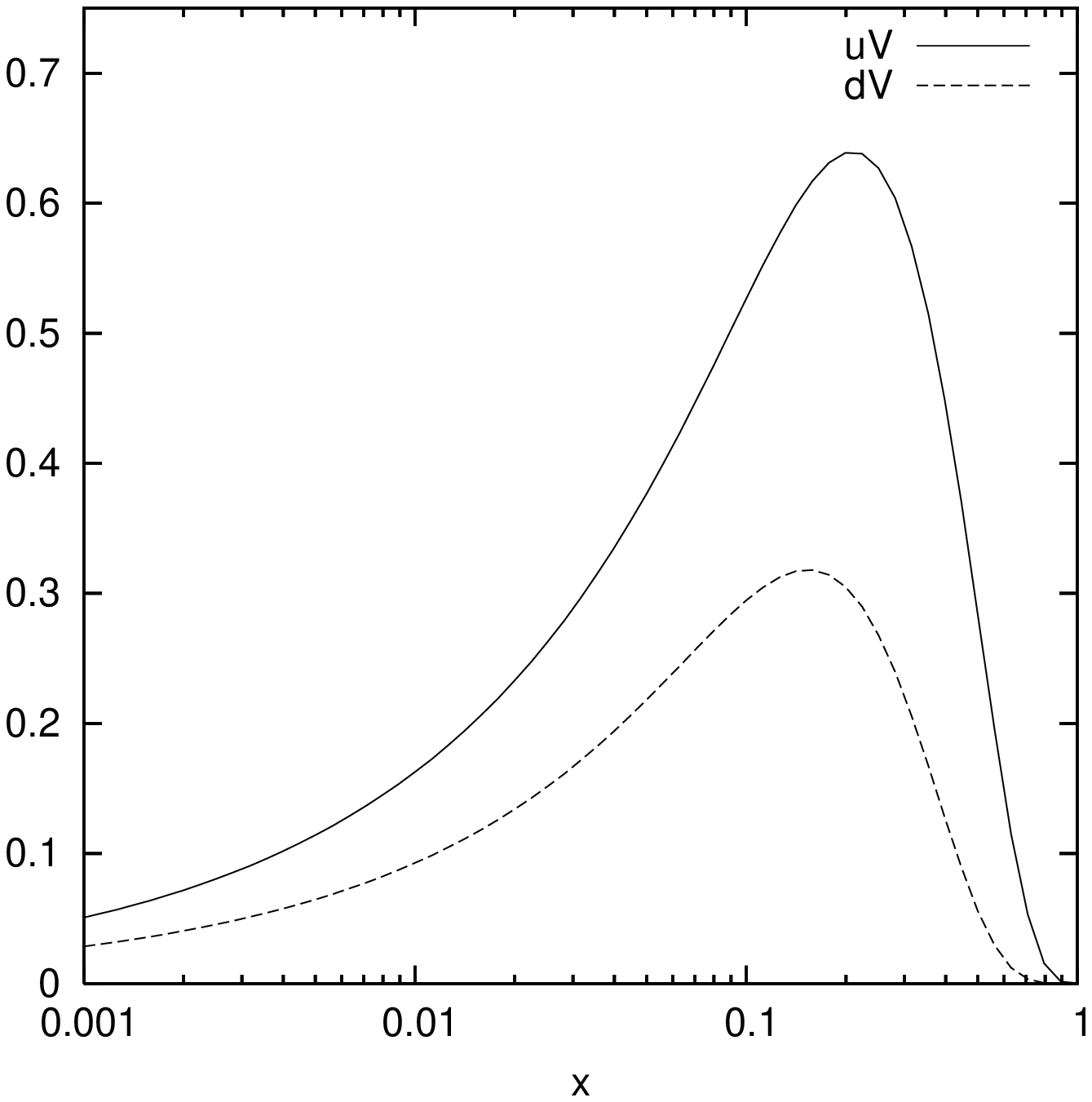}
\includegraphics[height=.32\textwidth,angle=270]{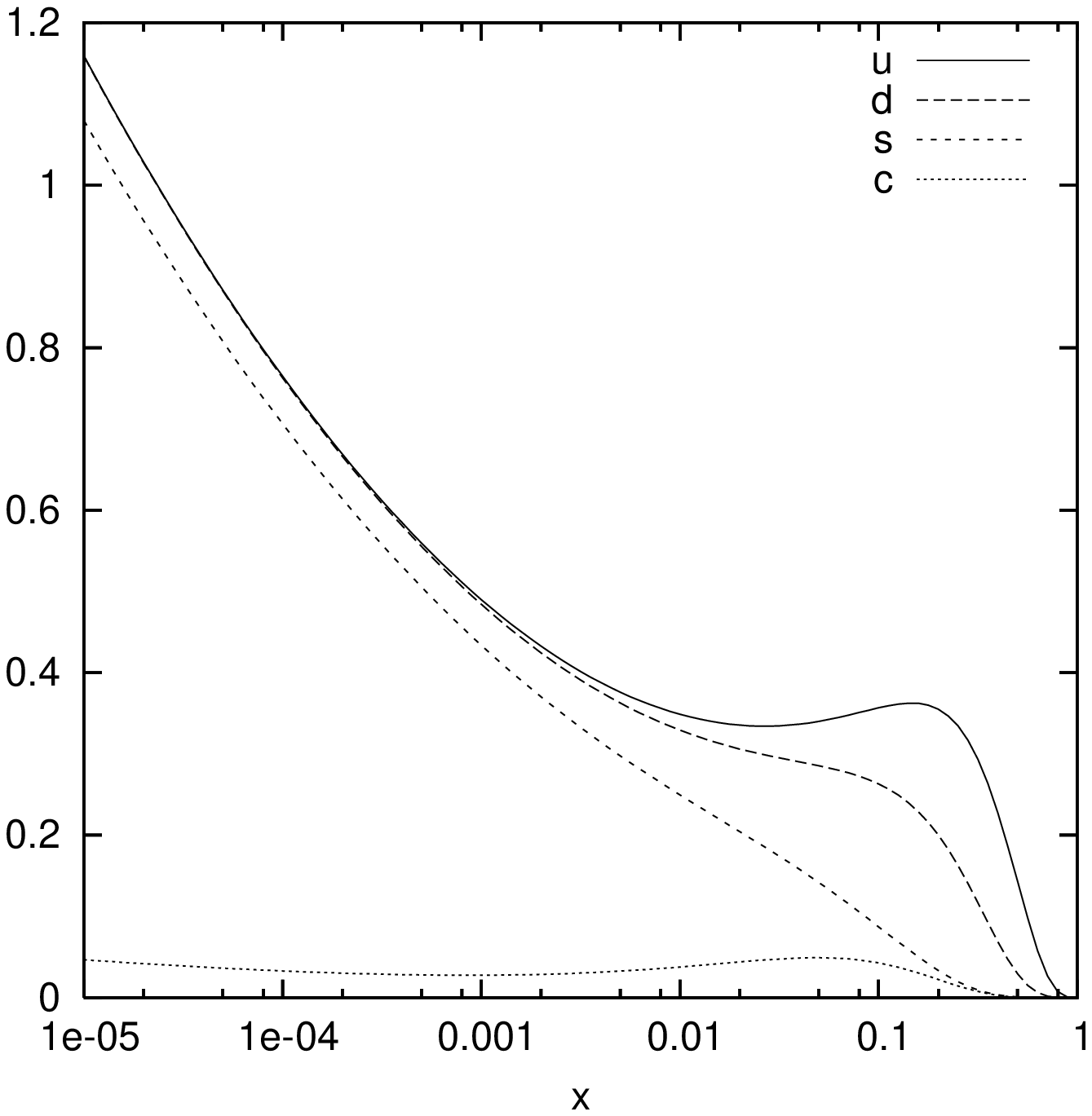}
\includegraphics[height=.32\textwidth,angle=270]{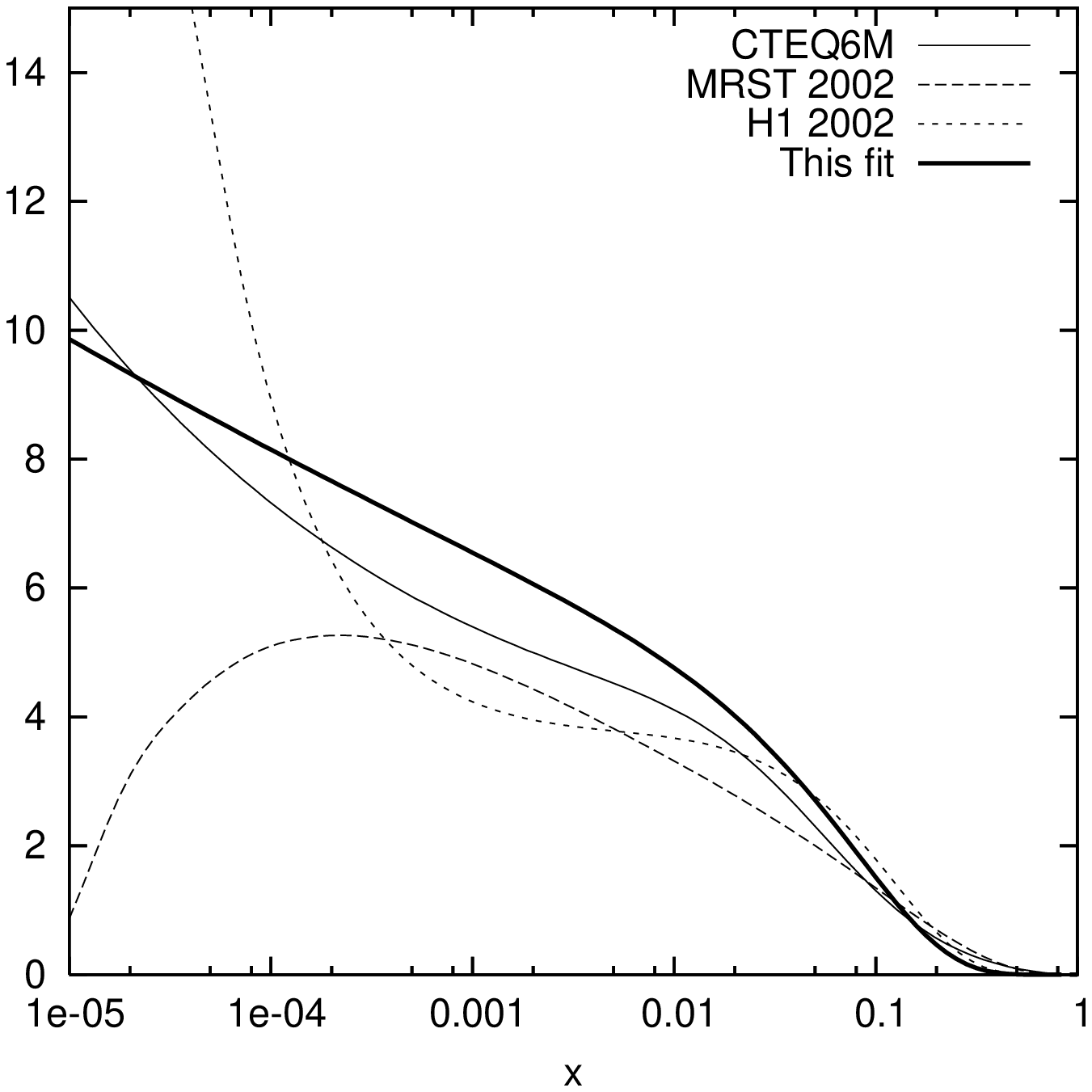}
\caption{Valence quark, sea quark and gluon distributions for the NLO global QCD fit.}
\label{fig:quarks}
\end{figure}

\begin{figure}
\includegraphics[height=.99\textwidth,angle=270]{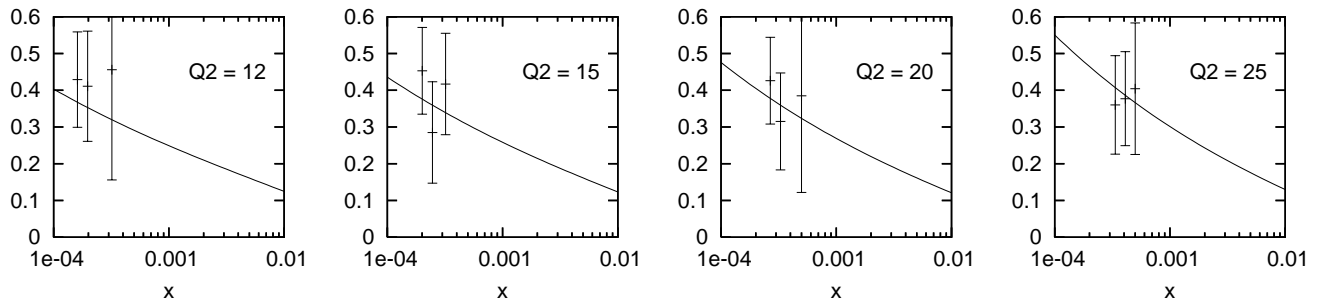}
\caption{Predictions of our fit for the longitudinal structure function $F_L$.}
\label{fig:fl}
\end{figure}

\begin{figure}
\includegraphics[height=.99\textwidth,angle=270]{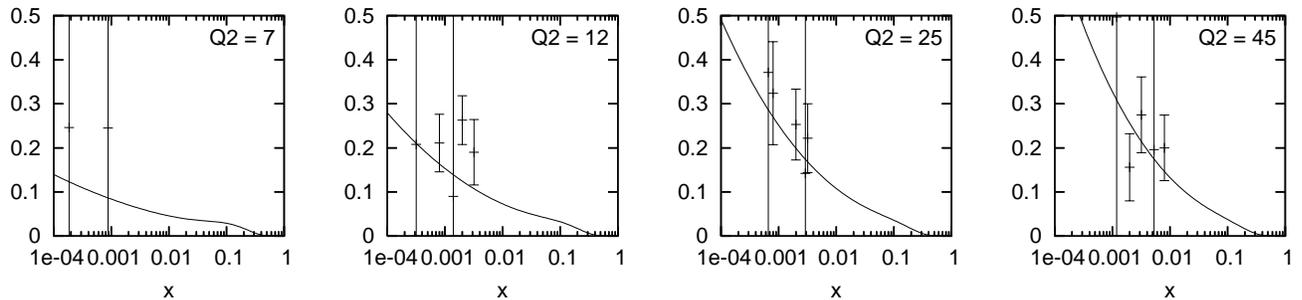}
\caption{Predictions of our fit for the charm structure function $F^2_c$.}
\label{fig:fc}
\end{figure}

Let us show how to solve this problem \cite{Soyez:2004ni}. We shall perform a global QCD fit starting at $Q_0^2=5$ GeV$^2$ with initial parton distributions compatible with Regge theory. In other words, one need to parametrise the initial parton distribution in agreement with $S$-matrix theory. In order to obtain a complete PDF set, one need to parametrise the valence quarks $u_v$ and $d_V$, the sea quarks $u_s$, $d_s$, $s_s$ and $c_s$, and the gluon density $g$. These expressions have to reproduce the pomeron and reggeon exchanges. Inserting a power of $1-x$ in order to ensure that the parton distributions go to 0 as $x$ goes to 1, this gives
\begin{eqnarray}
xu_V & = & \frac{2}{N_u^*} x^\eta(1+\gamma_u x)(1-x)^{b_u},\nonumber\\
xd_V & = & \frac{1}{N_d^*} x^\eta(1+\gamma_d x)(1-x)^{b_d},\nonumber\\
xu_s & = & \left[A\log^2(1/x)+B\log(1/x)+C+D_u x^\eta \right](1-x)^b,\nonumber\\
xd_s & = & \left[A\log^2(1/x)+B\log(1/x)+C+D_d x^\eta \right](1-x)^b,\label{eq:init}\\
xs_s & = & N_s\left[A\log^2(1/x)+B\log(1/x)+C+D_s x^\eta \right](1-x)^b,\nonumber\\
xc_s & = & N_c\left[A\log^2(1/x)+B\log(1/x)+C+D_s x^\eta \right](1-x)^b,\nonumber\\
xg   & = & \left[A_g\log^2(1/x)+B_g\log(1/x)+C_g^*\right](1-x)^{b+1},\nonumber
\end{eqnarray}
In these distributions, the $\log^2(1/x)$, the $\log(1/x)$ and the constant terms correspond to a triple-pole pomeron exchange while the fourth one, with $\eta = 0.4$ is a $f$-reggeon exchange. To build the initial conditions \eqref{eq:init}, we have applied the following physical arguments
\begin{itemize}
\item The pomeron is a flavour-singlet object. It therefore decouples from valence quarks and couples in the same way to all sea quarks.
\item The reggeon is expected to represent quark exchanges. We assume that it is not coupled to gluons.
\item The power of $1-x$ defining the large-$x$ behaviour of the parton distributions is taken to be the same for all sea quarks. The exponent in the gluon distribution appears from a study of the DGLAP equation at large $x$ \cite{lopez}.
\item The mass effects has been introduced through the normalisation factors $N_s$ and $N_c$. Due to the small strange mass, we expect $N_s \approx 1$. $N_c$ is non-zero for initial scales $Q_0^2$ larger than $m_c^2 = 2$ GeV$^2$. Similarly, during the DGLAP evolution, the $b$ quarks will be switched on at $Q^2 = m_b^2 = 20.25$ GeV$^2$.
\item Quark number conservation fixes $N_u$ and $N_d$. The momentum sum rule is used to constrain $C_g$.
\end{itemize}

We have performed a global QCD fit with this initial condition at leading and next-to-leading order (in the $\overline{MS}$ scheme). We have included the data for the proton \cite{H1,ZEUS,BCDMS,E665,NMC} and deuteron \cite{E665,BCDMSd,NMCd} structure functions, the ratio $F_2^n/F_2^p$ \cite{NMCd} and the neutrino structure functions \cite{CCFR2,CCFR3}. The initial scale $Q_0^2$ has been set to 5 GeV$^2$ and we have imposed $W^2 \ge 12.5$ GeV$^2$ in order to cut the region where we expect higher-twist effects. We obtain, as presented in Table \ref{tab:chi2}, a chi-square per data point slightly larger than 1. This is at least as good as the results reached by standard global fits.

In Fig. \ref{fig:quarks} we show the valence and sea quark distributions. We also present the gluons obtained from our NLO fit compared with some distributions taken from various standard PDF sets at $Q^2=5$ GeV$^2$. Our distributions are compatible with the standard PDF sets.

Finally, one can consider predictions for the charm structure function $F_2^c$ or for the longitudinal structure function $F_L$ \cite{FL}. These are sensitive to the gluon distribution and, in addition, $F_2^c$ \cite{Fc} is a good test for the charm distribution. Our predictions are compatible with the data, as presented in Figs. \ref{fig:fl} and \ref{fig:fc}. We also obtain a good description of the Tevatron Drell-Yan data with a $K$-factor between 1.3 and 1.4.

\section{Small-$Q^2$ description}

\begin{figure}
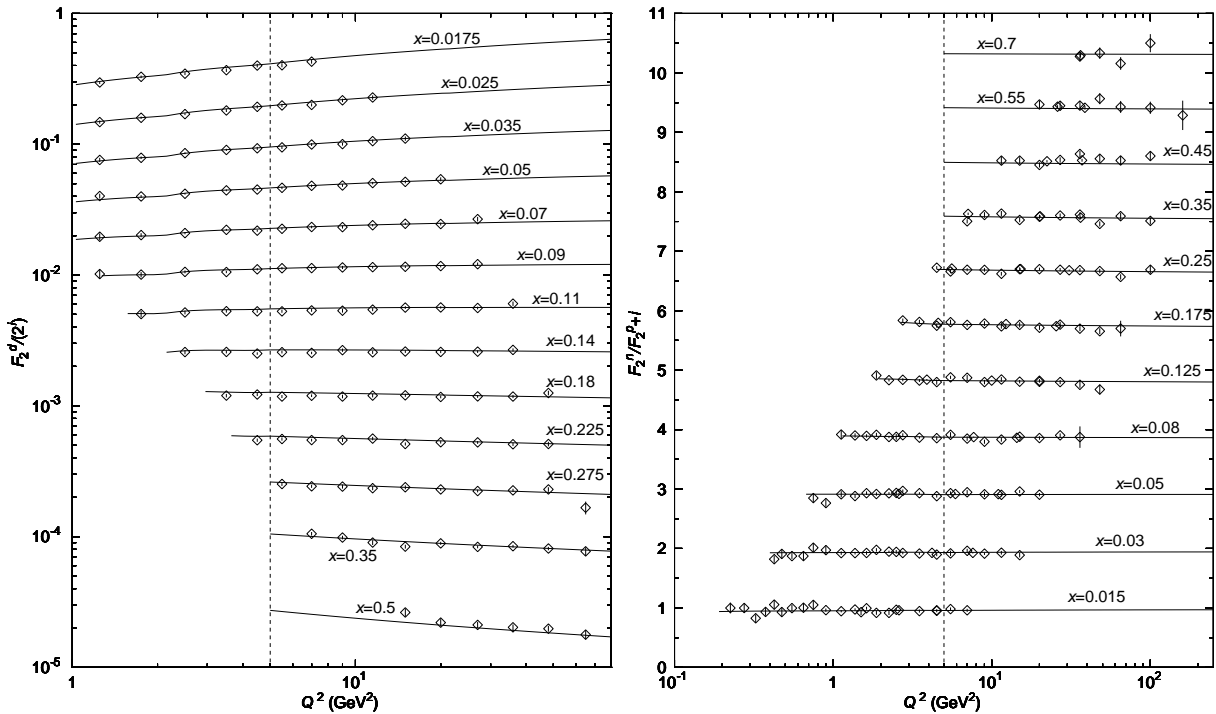

  \includegraphics[height=0.4\textheight]{f2d-nmc.ps}
  \includegraphics[height=0.4\textheight]{f2np.ps}
  \caption{Result of our fit for the NMC data for $F_2^d$ (left) and $F_2^n/F_2^p$ (right). In each plot, the high-$Q^2$ part is described by DGLAP evolution and the low-$Q^2$ data by Regge theory.}
  \label{fig:nmc}
\end{figure}

One of the main advantages of restoring relevant analytic expressions for the initial distribution is that one can use \cite{Soyez:2004pz} Regge-theory techniques to describe the data for $0\le Q^2 \le Q_0^2$. We need to use the parametrisation \eqref{eq:init} and let the coefficients of the fit being $Q^2$-dependent. For the $Q^2$ dependence, we have adopted for these dependences, the usual parametrisation
\[
\phi(Q^2) = \phi(Q_0^2) \frac{Q^2}{Q_0^2}\left(\frac{Q_0^2+Q_\phi^2}{Q^2+Q_\phi^2}\right)^{\varepsilon_\phi},
\]
where we have imposed matching with the parameters obtained from the global fit at $Q^2=Q_0^2$. With a few simplifications, {\em e.g.}, $Q^2_{D_u}=Q^2_{D_d}=Q^2_{D_s}=Q^2_{D_c}$, it gives a total of 13 parameters.

At leading order, it is sufficient to consider extensions of the quark distributions. Due to the convolution with coefficient functions at NLO, one also needs small-$Q^2$ expressions for the gluon distribution and for the strong coupling constant $\alpha_s$. Since we expect our results to be relatively independent of this parametrisation, we have simply considered that the gluon density scales with $Q^2$ ($g(x,Q^2) = Q^2/Q_0^2\, g(x,Q_0^2)$) and that the running coupling stays constant for $Q^2\le m_c^2$.

The $\chi^2$ resulting from this fit is presented in Table \ref{tab:chi2} together with the global QCD fit. {\em We see that at NLO, one finally obtains a description of the DIS data, at all values of $Q^2$, for $W^2\ge12.5$ GeV$^2$ with a chi-square of 1 per data point.}

To illustrate this, we have shown in Fig. \ref{fig:nmc} the final result of our fit compared with the $F_2^d$ and $F_2^n/F_2^p$ measures from NMC.

\section{Regge theory at high-$Q^2$}

\begin{figure}
\includegraphics{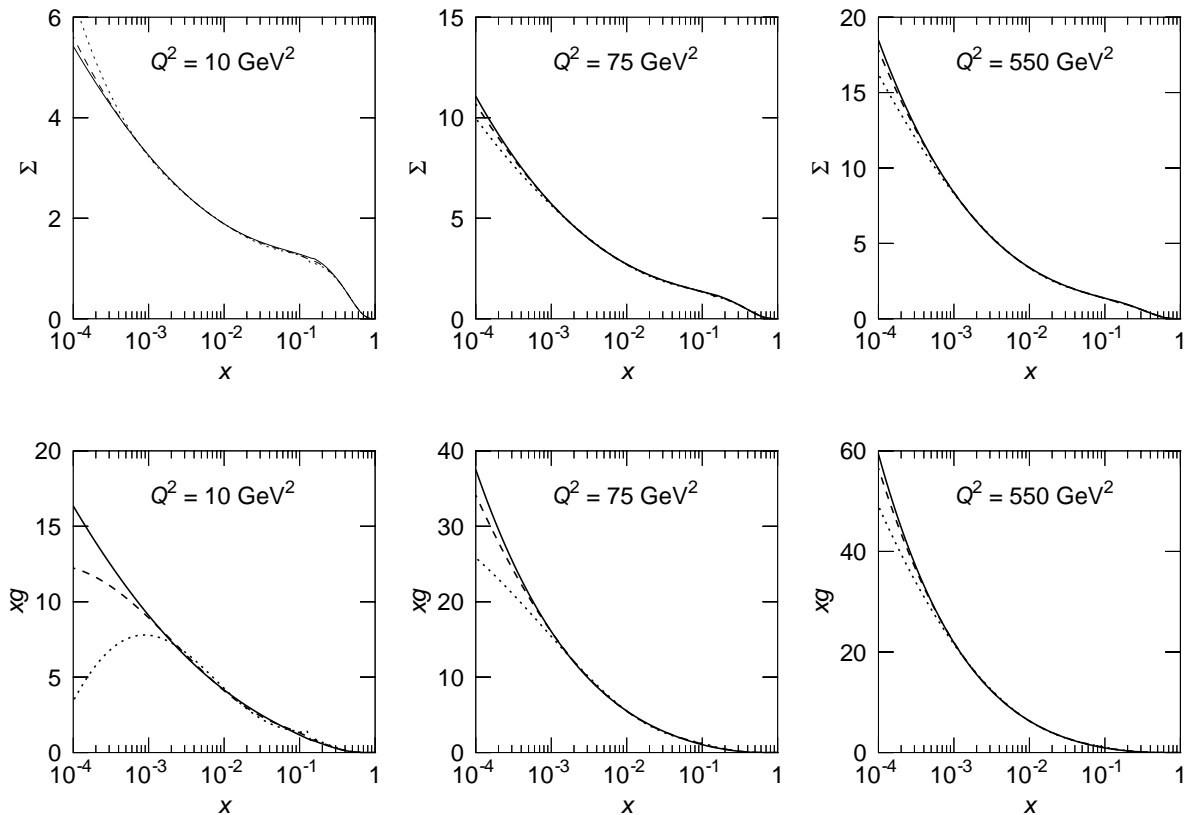}
\caption{Sea quark ($\Sigma$) and gluon ($xg$) distributions at high $Q^2$. The solid (resp. dashed, dotted) line corresponds to the fit starting at $Q_0^2 = 10$ GeV$^2$ (resp. 75, 550 GeV$^2$). For example, the solid curve has a triple-pole pomeron behaviour at $Q^2=10$ GeV$^2$ and a essential singularity at other $Q^2$ values.}
\label{fig:pdfqg}
\end{figure}

We have shown that it is possible to have a description of the DIS data based on Regge theory at small $Q^2$ and on DGLAP evolution at large $Q^2$. However, it is well-known that DGLAP evolution generates an essential singularity at $j=1$, which seems in disagreement with Regge theory.

It is proven that, in the small-$x$ region, additional resummation needs to be done if we want to recover a Regge-like high-energy behaviour. Our statement is that, even if their analytic structure are not physical, the DGLAP-obtained parton densities have to be treated as numerical approximations. Our aim is to show that, if we assume that DGLAP evolution approximates a triple-pole pomeron, we can extract the residues of the pomeron at large $Q^2$ from the DGLAP equation. In order to find the residues we may therefore adopt the following strategy \cite{Soyez:2003sr}:
\begin{enumerate}
\item\label{s21} choose an initial scale $Q_0^2$ at which one searches for the Regge residues,
\item\label{s22} choose a value for the parameters in the initial distribution,
\item\label{s23} compute the parton distributions for $Q_0^2 \le Q^2  \le Q_{\text{max}}^2$ using forward DGLAP evolution and for $Q_{\text{min}}^2 \le Q^2  \le Q_0^2$ using backward DGLAP evolution,
\item\label{s24} repeat \ref{s22} and \ref{s23} until the value of the parameters reproducing the $F_2$ data for $Q^2>Q_{\text{min}}^2$ and $x\le 0.15$ is found.
\item This gives the residues at the scale $Q_0^2$ and steps \ref{s21} to \ref{s24} are repeated in order to obtain the residues at all $Q^2$ values.
\end{enumerate}

This technique allows us to obtain the residues of a Regge fit directly from QCD evolution, without having to postulate an analytic expression for the $Q^2$ dependence. It is also motivated by the fact that physics is expected to be independent on the choice of $Q_0^2$.

Applying this method between 10 and 1000 GeV$^2$ and for $x\le 0.15$, we have obtained the $Q^2$-dependent residues of the pomeron giving a fit to $F_2^p$ with $\chi^2/nop=1.02$. One may then check that the DGLAP result is a correct numerical approximation to a triple-pole pomeron model. In Fig. \ref{fig:pdfqg}, we show the results of the DGLAP fit for 3 different choices of the initial scale. As seen from the upper part of the plot, in the case of the sea quark distribution, it is impossible to distinguish between the triple-pole pomeron behaviour ($Q^2=Q_0^2$) and the curves with a DGLAP-generated essential singularity ($Q^2\neq Q_0^2$). This validates our initial argument.

The situation is not as clear for the gluon distributions (lower part of Fig. \ref{fig:pdfqg}). Since each of these fits reproduces equally well the $F_2^p$ data, the differences between them have to be considered as errors on the gluon distribution. We thus predict large errors on the gluon density at small $Q^2$ and small $x$, which may be of prime importance for LHC physics.

\section{Conclusions}

We have shown that it was possible to use Regge theory to constrain the parametrisation of the initial parton distributions in a global QCD fit. We obtain a standard PDF set at LO and NLO which reproduces the usual features of global QCD fits with the advantage of being consistent with analytic $S$-matrix theory. 

This allows us to extend the fit in the small-$Q^2$ region giving a complete description of the data, for $W^2\ge 12.5$ GeV$^2$, over the whole $Q^2$ range.

Finally, we have discussed the compatibility of Regge theory and DGLAP evolution at large $Q^2$. We show that one can use the DGLAP equation to extract the residues of the triple-pole pomeron and that this predicts large errors on the gluon distribution at small $x$ and small $Q^2$.

In the future, it might be interesting to repeat this Regge-constrained global QCD fit with other Regge models and to combine the low- and high-$Q^2$ fits in order to allow for a determination of the initial scale for DGLAP evolution.

The high-$Q^2$ study also invites one to look for QCD corrections to the DGLAP equation which stabilises the Regge behaviour.

%%%%%%%%%%%%%%%%%%%%%%%%%%%%%%%%%%%%%%%%%%%%%%%%
%% BACKMATTER
%%%%%%%%%%%%%%%%%%%%%%%%%%%%%%%%%%%%%%%%%%%%%%%%

\begin{acknowledgments}
G.S. is funded by the National Funds for Scientific Research (FNRS, Belgium)
\end{acknowledgments}

\end{document}